\documentclass{article}

\usepackage{stwol}
\usepackage{cite}
\usepackage{subeqn}
\usepackage[dvips]{graphicx}
\usepackage{ifthen}

\newcommand{\prenumber}{1}	% 1=on, 0=off
\def\openone{\leavevmode\hbox{\small1\kern-3.8pt\normalsize1}}%
\def\slash#1{\setbox0=\hbox{$#1$}#1\hskip-\wd0\dimen0=5pt\advance
       \dimen0 by-\ht0\advance\dimen0 by\dp0\lower0.5\dimen0\hbox
         to\wd0{\hss\sl/\/\hss}}

\def\CPbar{\setbox0=\hbox{\textrm{CP}}\setbox1=\hbox{\textsl{\LARGE/}}
	\kern-0.3em\box0\kern-1em\lower0.2ex\box1\kern0.3em}

\newlength{\miniwidth}
\newlength{\miniwidthplot}
\newlength{\minicolumn}
\newlength{\tseparation}
\newlength{\bseparation}
\newenvironment{nfigure}
	{\begin{figure}[htbp]\hrule\vspace{\tseparation}\par}
	{\vspace{\bseparation}\par \hrule \end{figure}}
\newenvironment{ntable}
	{\begin{table}[htbp]\hrule\vspace{\tseparation}\par}
	{\vspace{\bseparation}\par \hrule \end{table}}
\newcommand{\ncaption}[1]{\caption{\slshape #1}}

\newcommand{\ov}{\overline}
\newcommand{\nn}{\nonumber}
\newcommand{\diff}[1]{\frac{d}{d #1}}

\newcommand{\pole}{\mathrm{pole}}
\newcommand{\eps}{\varepsilon}
\newcommand{\Lagr}{{\mathcal{L}}}
\newcommand{\eff}{{\mathrm{eff}}}
\newcommand{\eq}[1]{(\ref{#1})}
\newcommand{\fig}[1]{Fig.~\ref{#1}}
\newcommand{\tab}[1]{Table~\ref{#1}}

\newcommand{\imag}{{\mathrm{Im}\,}}

\newcommand{\mw}{M_W^2}
\newcommand{\as}{\alpha_s}
\newcommand{\hc}{\mathrm{h.c.}}

\newcommand{\gev}{\,\mathrm{GeV}}
\newcommand{\mev}{\,\mathrm{MeV}}
\newcommand{\epsK}{\ensuremath{\eps_\mathrm{K}}}
\newcommand{\DmK}{\ensuremath{\Delta m_\mathrm{K}}}
\newcommand{\bk}{\ensuremath{B_\mathrm{K}}}
\newcommand{\fk}{\ensuremath{f_\mathrm{K}}}
\newcommand{\mk}{\ensuremath{m_\mathrm{K}}}
\newcommand{\DmBd}{\ensuremath{\Delta m_\mathrm{B_d}}}
\newcommand{\DmBs}{\ensuremath{\Delta m_\mathrm{B_s}}}
\newcommand{\bbd}{\ensuremath{B_\mathrm{B_d}}}
\newcommand{\bbs}{\ensuremath{B_\mathrm{B_s}}}
\newcommand{\fbd}{\ensuremath{f_\mathrm{B_d}}}
\newcommand{\fbs}{\ensuremath{f_\mathrm{B_s}}}
\newcommand{\mb}{\ensuremath{m_\mathrm{B}}}
\newcommand{\mbd}{\ensuremath{m_\mathrm{B_d}}}
\newcommand{\mbs}{\ensuremath{m_\mathrm{B_s}}}
\newcommand{\laMSb}{\ensuremath{
	\Lambda_{\scriptscriptstyle\overline{\mathrm{MS}}}}}
\newcommand{\laQCD}{\ensuremath{\Lambda_{\scriptscriptstyle\mathrm{QCD}}}}
\newcommand{\gf}{G_\mathrm{\!\scriptscriptstyle F}}
\newcommand{\mmmix}[1]{\ensuremath{\mathrm{#1\!-\!\ov{#1}}\,}-mixing\ }

\newcommand{\bbm}{\mmmix{B^0}}
\newcommand{\bbmd}{\mmmix{B_d^0}}
\newcommand{\bbms}{\mmmix{B_s^0}}

\newcommand{\dstwo}{\ensuremath{\mathrm{|\Delta S| \!=\!2}}}
\newcommand{\dsone}{\ensuremath{\mathrm{|\Delta S| \!=\!1}}}

\newcommand{\msb}{\ensuremath{\ov{\textrm{MS}}}}
\newcommand{\dmu}{\mu\!\diff{\mu}}

\newcommand{\vckm}{\ensuremath{V_\mathrm{\scriptscriptstyle CKM}}}

\newcommand{\oll}{\ensuremath{\tilde{Q}_{S2}}}

\newcommand{\errorpm}[2]{
\raisebox{-0.5ex}{\shortstack[l]{$\scriptstyle+#1$\\$\scriptstyle-#2$}}
}
\newcommand{\pr}{Phys.\ Rev.\ }
\newcommand{\prd}{\pr D}
\newcommand{\np}{Nucl.\ Phys.\ }
\newcommand{\npb}{\np B}

\newcommand{\prp}{preprint }

\newcommand{\hn}{S.~Herrlich and U.~Nierste}
\newcommand{\here}{these proceedings.}

\newcommand{\pls}{plenary sessions, \here}

\begin{document}

% titlepage if preprint
\ifthenelse{\prenumber=1}{
\begin{titlepage}
\renewcommand{\thefootnote}{\fnsymbol{footnote}}
\setcounter{page}{0}
\begin{flushright}
\textsf{
hep-ph/9609376\\
DESY 96-190 \\
September 1996
}
\end{flushright}
\vspace{5em}\par
\begin{center}
\textbf{
{\Large The Complete \dstwo\ Hamiltonian in the Next-To-Leading Order}\\[2em]
{\large and its}\\[2em]
{\Large Phenomenological Implications\footnote{Report on work done in
collaboration with U.\ Nierste.  To appear in the Proceedings of
ICHEP'96, 24.--31. Aug. 1996, Warsaw, Poland} }
}
\vspace{5em} \par
{\large S.\ Herrlich\footnote{e-mail
\texttt{Stefan.Herrlich@feynman.t30.physik.tu-muenchen.de}}}
\vspace{2em}
\par
\textit{DESY-IfH Zeuthen Platanenallee 6, D-15738 Zeuthen, Germany}\\
\end{center}
\setcounter{footnote}{0} % reset the counter
\end{titlepage}
}{} % end of titlepage

% now the head of the article...
\title{
THE COMPLETE $\dstwo$ HAMILTONIAN IN THE NEXT-TO-LEADING ORDER
AND ITS PHENOMENOLOGICAL IMPLICATIONS
}

\author{STEFAN HERRLICH}

\address{DESY-IfH Zeuthen, Platanenallee 6, D-15738 Zeuthen, Germany}

\twocolumn[\maketitle\abstracts{ We briefly sketch the calculation of
the effective low-energy \dstwo-hamiltonian in the next-to-leading
order of renormalization group improved perturbation theory.  The
result for the coefficient $\eta_3^\star$ is discussed.  Further we
present a 1996 update of our phenomenological analysis of the
unitarity triangle where we include the information available on
\bbm.  }]

\section{The \dstwo-Hamiltonian}

Here we briefly report on the \dstwo-hamiltonian, the calculation of
its next-to-leading order (NLO) QCD corrections and on the numerical
results.  For the details we refer to \cite{hn}.

\subsection{The low-energy \dstwo-Hamiltonian}

The effective low-energy hamiltonian inducing the \dstwo-transition
reads:
\begin{eqnarray}
H^\dstwo_\eff
&=&
\frac{\gf^2}{16\pi^2}\mw
\bigl[
\lambda_c^2 \eta_1^\star S\!\left(x_c^\star\right)
+\lambda_t^2 \eta_2^\star S\!\left(x_t^\star\right)
\nn\\&&
+2 \lambda_c \lambda_t \eta_3^\star S\!\left(x_c^\star,x_t^\star\right)
\bigr]
b\!\left(\mu\right) \oll\!\left(\mu\right)
\nn\\&&
+\hc
\label{Heff}
\end{eqnarray}
Here $\gf$ denotes Fermi's constant, $M_W$ is the W boson mass,
$\lambda_j = V_{jd} V_{js}^\ast, j=c,t$ comprises the CKM-factors, and
$\oll$ is the local dimension-six \dstwo\ four-quark operator
\begin{equation}
\oll =
\left[\bar{s}\gamma_\mu\left(1-\gamma_5\right)d\right]
\left[\bar{s}\gamma^\mu\left(1-\gamma_5\right)d\right]
\label{DefOll}
\end{equation}
The $x_q^\star = {m_q^\star}^2/\mw$, $q=c,t$ encode the running
{\msb}-quark masses $m_q^\star = m_q\!\left(m_q\right)$.  In writing
\eq{Heff} we have used the GIM mechanism
$\lambda_u+\lambda_c+\lambda_t=0$ to eliminate $\lambda_u$, further we
have set $m_u=0$.  The Inami-Lim functions $S\!\left(x\right)$,
$S\!\left(x,y\right)$ contain the quark mass dependence of the
{\dstwo}-transition in the absence of QCD.  They are obtained by
evaluating the box-diagrams displayed in {\fig{fig:full-lo}}.
\begin{nfigure}
\begin{minipage}[t]{\minicolumn}
\includegraphics[clip,width=0.8\minicolumn]{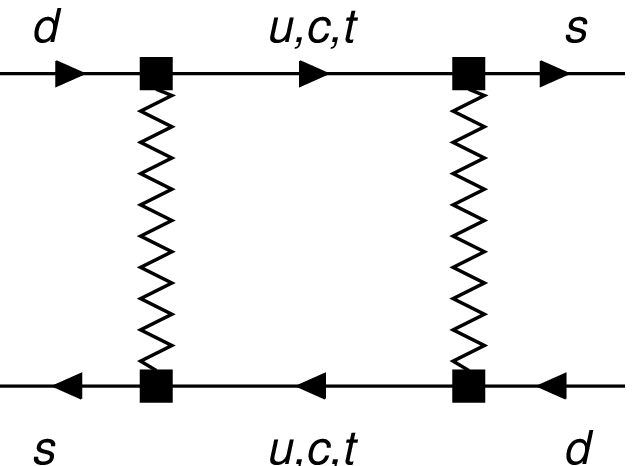}
\end{minipage}
\hfill
\begin{minipage}[b]{\minicolumn}
\ncaption{
The lowest order box-diagram mediating a \dstwo-transition.  The
zig-zag lines denote W-bosons or fictitious Higgs particles.}
\label{fig:full-lo}
\end{minipage}
\end{nfigure}

In \eq{Heff} the short-distance QCD corrections are comprised in the
coefficients $\eta_1$, $\eta_2$, $\eta_3$ with their explicit
dependence on the renormalization scale $\mu$ factored out in the
function $b\!\left(\mu\right)$.  The $\eta_i$ depend on the
\emph{definition} of the quark masses.  In \eq{Heff} they are
multiplied with $S$ containing the arguments $m_c^\star$ and
$m_t^\star$, therefore we marked them with a star.  In absence of QCD
corrections $\eta_i b\!\left(\mu\right)=1$.

For physical applications one needs to know the matrix-element of
$\oll$ \eq{DefOll}.  Usually it is parametrized as
\begin{equation}
\left\langle \overline{\mathrm{K^0}} \left| \oll\!\left(\mu\right)
\right| \mathrm{K^0} \right\rangle =
\frac{8}{3} \frac{\bk}{b\!\left(\mu\right)} \fk^2 \mk^2.
\label{DefBK}
\end{equation}
Here $\fk$ denotes the Kaon decay constant and $\bk$ encodes the
deviation of the matrix-element from the vacuum-insertion result.  The
latter quantity has to be calculated by non-perturbative methods.  In
physical observables the $b\!\left(\mu\right)$ present in \eq{DefBK}
and \eq{Heff} cancel to make them scale invariant.

The first complete determination of the coefficients $\eta_i$,
$i=1,2,3$ in the leading order (LO) is due to Gilman and Wise
\cite{gw}.  However, the LO expressions are strongly dependent on the
factorization scales at which one integrates out heavy particles.
Further the questions about the \emph{definition} of the quark masses
and the QCD scale parameter \laQCD\ to be used in \eq{Heff} remain
unanswered.  Finally, the higher order corrections can be sizeable and
therefore phenomenologically important.

To overcome these limitations one has to go to the NLO.  This program
has been started with the calculation of $\eta_2^\star$ in \cite{bjw}.
Then Nierste and myself completed it with $\eta_1^\star$ \cite{hn2}
and $\eta_3^\star$ \cite{hn,hn1}.

We have summarized the result of the three $\eta_i^\star$'s in
{\tab{tab:result}}.
\begin{ntable}
\ncaption{The numerical result for the three $\eta_i^\star$ using
$\as\!\left(M_Z\right)=0.117$, $m_c^\star=1.3\gev$,
$m_t^\star=167\gev$ as the input parameters.  The error of the NLO
result stems from scale variations.}
\label{tab:result}
\begin{tabular}{l@{\hspace{2em}}*{3}{l}}
 & $\eta_1^\star$ & $\eta_2^\star$ & $\eta_3^\star$ \\
\hline
LO&
$\approx$ 0.74 &
$\approx$ 0.59 &
$\approx$ 0.37
\\
NLO&
1.31\errorpm{0.25}{0.22} &
0.57\errorpm{0.01}{0.01} &
0.47\errorpm{0.03}{0.04}
\end{tabular}
\end{ntable}

\subsection{A short glance at the NLO calculation of $\eta_3^\star$}

Due to the presence of largely separated mass scales \eq{Heff}
develops large logarithms $\log x_c$, which spoil the applicability of
naive perturbation theory (PT).  Let us now shortly review the
procedure which allows us to sum them up to all orders in PT, finally
leading to the result presented in \tab{tab:result}.  The basic idea
is to construct a hierarchy of effective theories describing {\dsone}-
and {\dstwo}-transitions for low-energy processes.  The techniques
used for that purpose are Wilson's operator product expansion (OPE)
and the application of the renormalization group (RG).

At the factorization scale $\mu_{tW}=O\!\left(M_W,m_t\right)$ we
integrate out the W boson and the top quark from the full Standard
Model (SM) Lagrangian.  Strangeness changing transitions\footnote{In
general all flavour changing transitions} are now described by an
effective Lagrangian of the generic form
\begin{equation}
\Lagr^\dstwo_\eff =
-\frac{\gf}{\sqrt{2}}\vckm\!\sum_k C_k Q_k
-\frac{\gf^2}{2}\vckm\!\sum_l \tilde{C}_k \tilde{Q}_k.
\label{LeffAbove}
\end{equation}
The $\vckm$ comprise the relevant CKM factors.  The $Q_k$
($\tilde{Q}_l$) denote local operators mediating \dsone- (\dstwo-)
transitions, the $C_k$ ($\tilde{C}_l$) are the corresponding Wilson
coefficient functions which may simply be regarded as the coupling
constants of their operators.  The latter contain the short distance
(SD) dynamics of the transition while the long distance (LD) physics
is contained in the matrix-elements of the operators.

The \dsone-part of {\eq{LeffAbove}} contributes to \dstwo-transitions
via diagrams with double operator insertions like the ones displayed
in {\fig{fig:cc-cc-lo}}.
\begin{nfigure}
\includegraphics[clip,width=0.9\minicolumn]{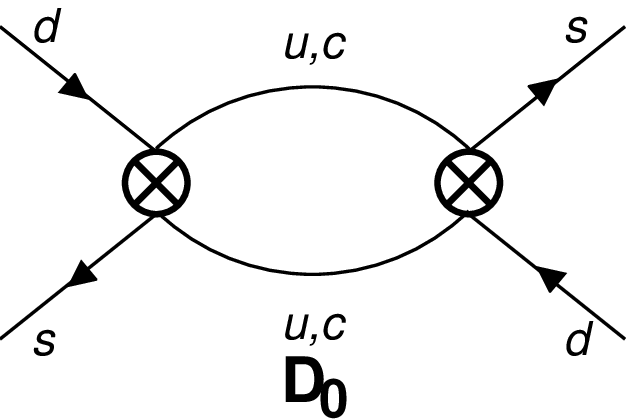}
\hfill
\includegraphics[clip,width=0.9\minicolumn]{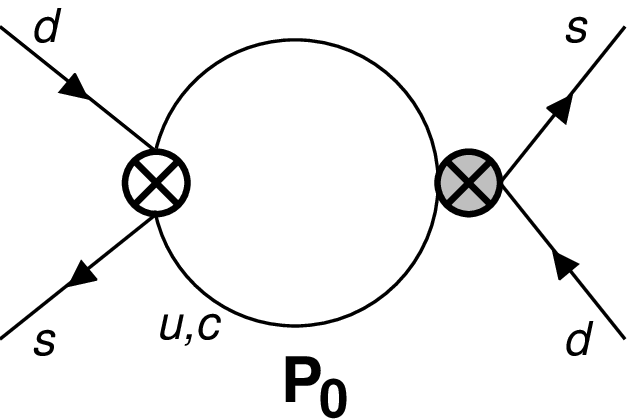}
\ncaption{
Two diagrams contributing to \dstwo-transitions in the effective
five- and four-quark theory.  The crosses denote insertions of
different species of local {\dsone}-operators.}
\label{fig:cc-cc-lo}
\end{nfigure}
The comparison of the Green's functions obtained from the full SM
lagrangian and the ones derived from \eq{LeffAbove} allows to fix the
values of the Wilson coefficients $C_k\!\left(\mu_{tW}\right)$ and
$\tilde{C}_l\!\left(\mu_{tW}\right)$.  The scale $\mu_{tW}$ being of
the order of $M_W$, $m_t$ ensures that there will be no large
logarithms in $C_k\!\left(\mu_{tW}\right)$ and
$\tilde{C}_l\!\left(\mu_{tW}\right)$, which therefore can be reliably
calculated in ordinary perturbation theory.

The next step is to evolve the Wilson coefficients
$C_k\!\left(\mu_{tW}\right)$, $\tilde{C}_l\!\left(\mu_{tW}\right)$
down to some scale $\mu_c = O\!\left(m_c\right)$, thereby summing up
the $\ln\left(\mu_c/\mu_{tW}\right)$ terms to all orders.\footnote{We
neglect the intermediate scale $\mu_b\!=\!O(m_b)$ for simplicity.}  To
do so, one needs to know the corresponding RG equations.  While the
scaling of the {\dsone}-coefficients is quite standard, the evolution
of the {\dstwo}-coefficients is modified due to the presence of
diagrams containing two insertions of {\dsone}-operators (see
{\fig{fig:cc-cc-lo}}).  From $\dmu\Lagr_\eff^\dstwo\!=\!0$ follows:
\begin{equation}
\dmu \tilde{C}_k\!\left(\mu\right) =
\tilde{\gamma}_{k'k} \tilde{C}_{k'}\!\left(\mu\right)
+\tilde{\gamma}_{ij,k} C_i\!\left(\mu\right) C_j\!\left(\mu\right).
\label{RGinhom}
\end{equation}
In addition to the usual homogeneous differential equation for
$\tilde{C}_l$ an inhomogenity has emerged.  The overall divergence of
diagrams with double insertions has been translated into an
{\textsl{anomalous dimension tensor}} $\tilde{\gamma}_{ij,k}$, which
is a straightforward generalization of the usual anomalous dimension
matrices $\gamma_{ij}$ ($\tilde{\gamma}_{ij}$).  The special structure
of the operator basis relevant for the calculation of $\eta_3$ allows
for a very compact solution of \eq{RGinhom} \cite{hn}.

Finally, at the factorization scale $\mu_c$ one has to integrate out
the charm-quark from the theory.  The effective three-flavour
lagrangian obtained in that way already resembles the structure of
{\eq{Heff}}, The only operator left over is $\oll$.  Double insertions
no longer contribute, they are suppressed with positive powers of
light quark masses.

We want to emphasize that throughout the calculation one has to be
very careful about the choice of the operator basis.  It contains
several sets of unphysical operators.  Certainly the most important
class of these operators are the so-called \emph{evanescent
operators}.  Their precise definition introduces a new kind of
scheme-dependence in intermediate results, e.g.\ anomalous dimensions
and matching conditions.  This scheme-dependence of course cancels in
physical observables.  Evanescent operators have been studied in great
detail in {\cite{hn3}}.

\subsection{Numerical Results for $\eta_3^\star$}

The numerical analysis shows $\eta_3^\star$ being only mildly
dependent on the physical input variables $m_c^\star$, $m_t^\star$ and
$\laMSb$ what allows us to treat $\eta_3^\star$ essentially as a
constant in phenomenological analyses.

More interesting is $\eta_3^\star$'s residual dependence on the
factorization scales $\mu_c$ and $\mu_{tW}$.  In principle
$\eta_3^\star$ should be independent of these scales, all residual
dependence is due to the truncation of the perturbation series.  We
may use this to determine something like a ``theoretical error''.

The situation is very nice with respect to the variation of
$\mu_{tW}$.  Here the inclusion of the NLO corrections reduces the
scale-dependence drastically compared to the LO.  For the interval
$M_W\leq\mu_{tW}\leq m_t$ we find a variation of less than 3\% in NLO
compared to the 12\% of the LO.

The dependence of $\eta_3^\star$ on $\mu_c$ has been reduced in NLO
compared to the LO analysis.  It is displayed in \fig{fig:e3s-muc}.
This variation is the source of the error of $\eta_3^\star$ quoted in
{\tab{tab:result}}.
\begin{nfigure}
\includegraphics[clip,width=0.65\miniwidth]{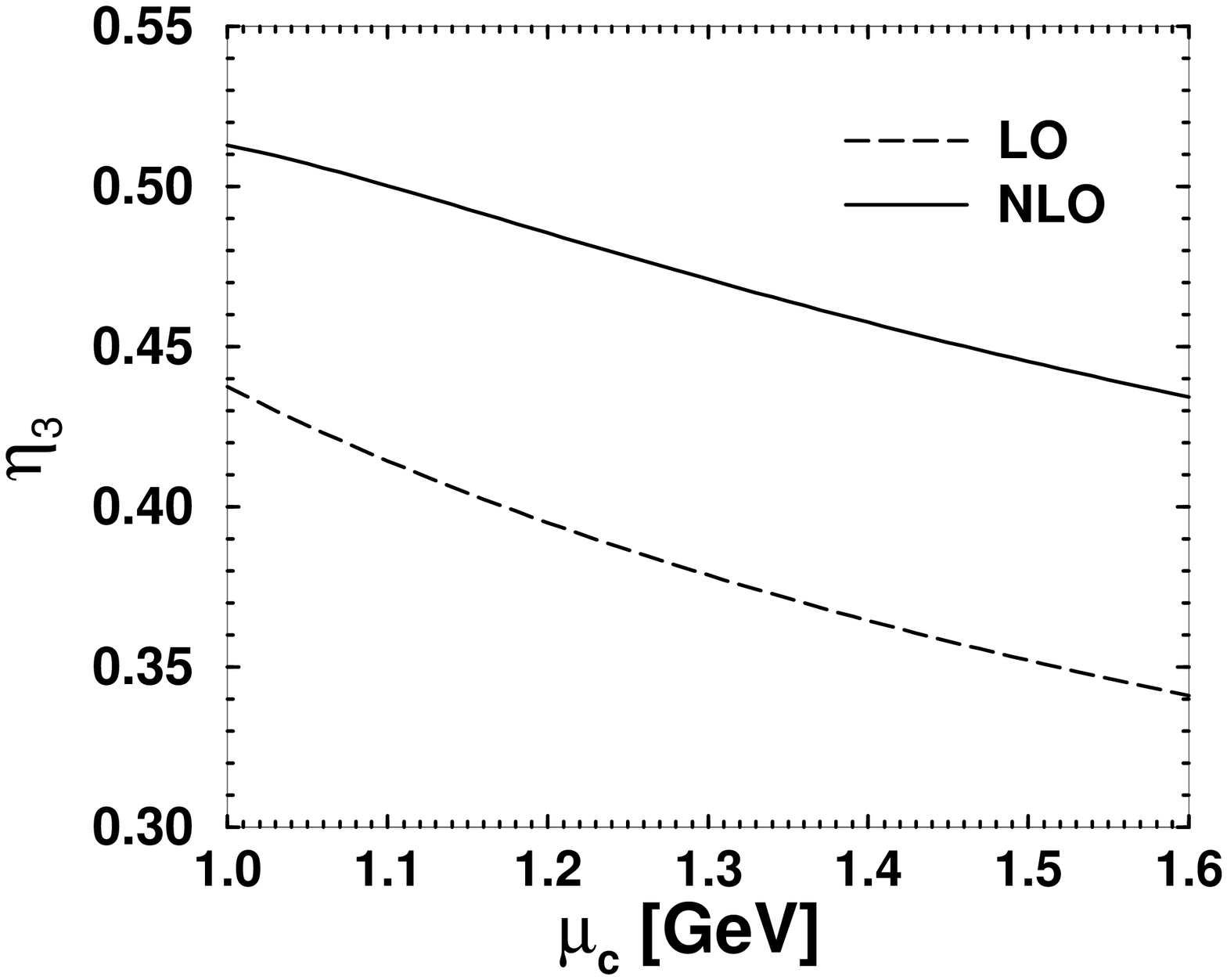}
\hfill
\parbox[b]{0.3\miniwidth}{
\ncaption{\sloppy
The variation of $\eta_3^\star$ with respect to the factorization
scale $\mu_c$, where the charm-quark gets integrated out.}}
\label{fig:e3s-muc}
\end{nfigure}

\section{The 1996 Phenomenology of $\left|\epsK\right|$}\label{sect:pheno}

The first phenomenological analysis using the full NLO result of the
{\dstwo}-hamiltonian has been done in \cite{hn1}.  Here we present a
1996 update.

\subsection{Input Parameters}

Let us first recall our knowledge of the CKM matrix as reported at
this conference \cite{ichep96-gibbons}:
\begin{subequations}
\begin{eqnarray}
\left|V_{cb}\right| &=& 0.040\pm0.003, \\
\left|V_{ub}/V_{cb}\right| &=& 0.08\pm0.02.
\end{eqnarray}
\end{subequations}

Fermilab now provides us with a very precise determination of
$m_t^\pole = 175\pm6 \gev${\cite{ichep96-tipton}} which translates
into the \msb-scheme as $m_t^\star = 167\pm6 \gev$.

There have been given more precise results on \bbmd and \bbms
{\cite{ichep96-gibbons}}:
\begin{subequations}
\begin{eqnarray}
\DmBd &=& \left(0.464\pm0.012\pm0.013\right) ps^{-1},\\
\DmBs &>& 9.2 ps^{-1}
\label{LimitDmBs}
\end{eqnarray}
\end{subequations}

We will further use some theoretical input:
\begin{subequations}
\begin{eqnarray}
\bk &=& 0.75\pm0.10\\
\fbd\sqrt{\bbd} &=& \left(200\pm40\right) \mev,\label{fbd}\\
\frac{\fbs\sqrt{\bbs}}{\fbd\sqrt{\bbd}} &=& 1.15\pm0.05.
\label{xisd}
\end{eqnarray}
\end{subequations}
{\eq{fbd}} and{\eq{xisd}} are from quenched lattice QCD, the latter
may go up by 10\% due to unquenching {\cite{ichep96-flynn}}.

The other input parameters we take as in \cite{hn1}.

\subsection{Results}

In extracting information about the still unknown elements of the CKM
matrix we still get the strongest restrictions from unitarity and
$\epsK$:
\begin{equation}
\left|\epsK\right| = \frac{1}{\sqrt{2}} \left[
	 \frac{\imag \left\langle \mathrm{K^0} \right|
	H^\dstwo \left| \overline{\mathrm{K^0}} \right\rangle}{\DmK}
	+\xi \right].
\label{epsKcond}
\end{equation}
Here $\xi$ denotes some small quantity related to direct \CPbar\
contributing about 3\% to $\epsK$.  The key input parameters entering
\eq{epsKcond} are $V_{cb}$, $|V_{ub}/V_{cb}|$, $m_t^\star$ and $\bk$

One may use \eq{epsKcond} to determine lower bounds on one of the four
key input parameters as functions of the other three.  In
{\fig{fig:vubcb-vcb}} the currently most interesting lower bound
curve which was invented in \cite{hn1} is displayed.
\begin{nfigure}
\includegraphics[clip,width=\miniwidth]{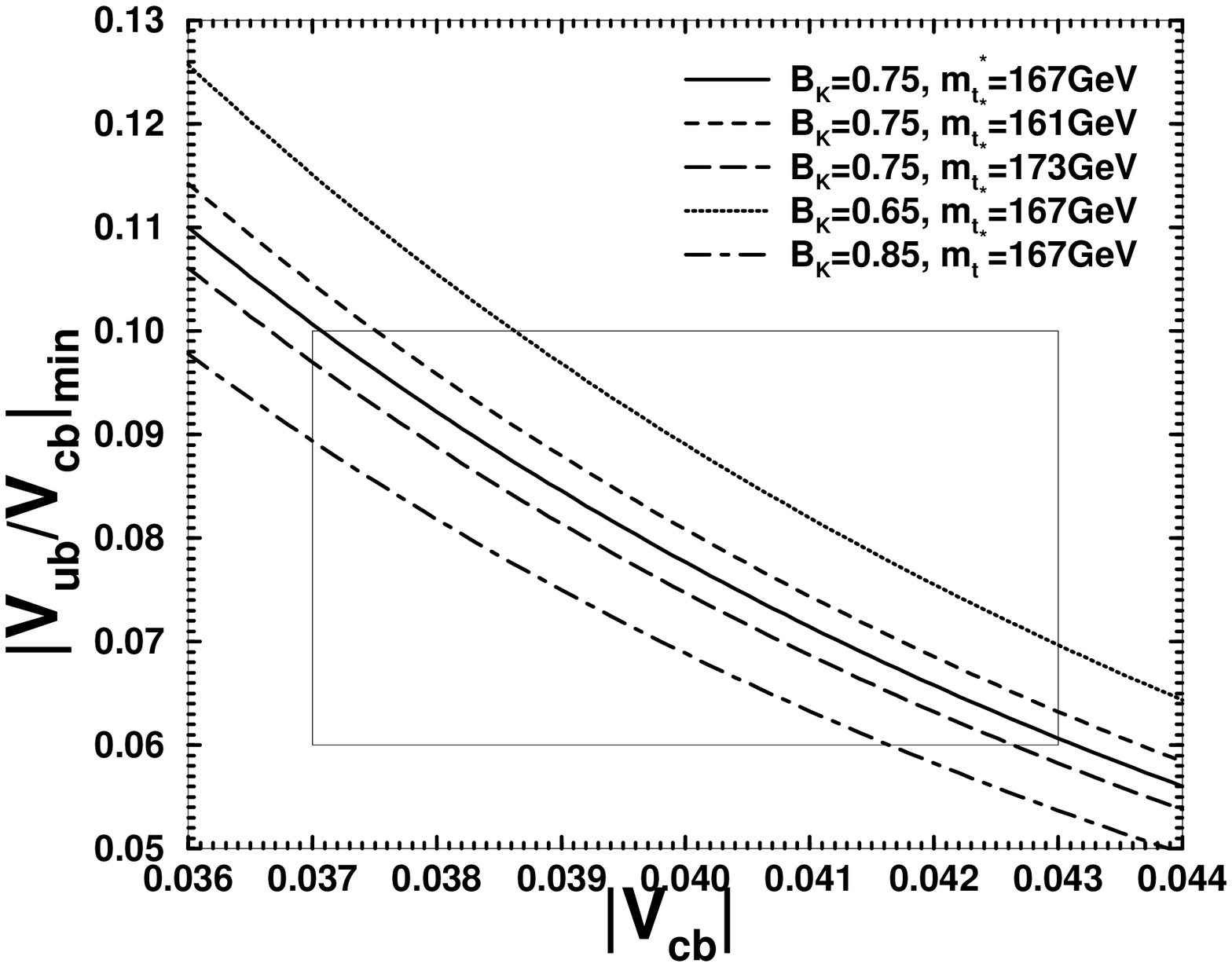}
\ncaption{ The lower-bound curves for $|V_{ub}/V_{cb}|$ as a function
of $V_{cb}$ for different values of the key input parameters
$m_t^\star$ and $\bk$.}
\label{fig:vubcb-vcb}
\end{nfigure}

Further we are interested in shape of the unitarity triangle, i.e.\
the allowed values of the top corner $\left(\bar\rho,\bar\eta\right)$
\begin{equation}
\bar\rho +i \bar\eta = - V_{ud} V_{ub}^* / V_{cd} V_{cb}^* .
\end{equation}
Here, in addition to \eq{epsKcond}, we take into account the constraint
from \bbmd
\begin{equation}
\DmBd = \left|V_{td}\right|^2 \left|V_{ts}\right|^2
\frac{\gf^2}{6\pi^2} \eta_\mathrm{B} \mb \bbd \fbd^2 \mw S\!\left(x_t\right)
\label{condBBmd}
\end{equation}
and \bbms
\begin{equation}
\DmBs = \DmBd \cdot \frac{\left|V_{ts}\right|^2}{\left|V_{td}\right|^2}
\cdot \frac{\mbd \fbd^2 \bbd}{\mbs \fbs^2 \bbs} .
\label{condBBms}
\end{equation}
The allowed region for $\left(\bar\rho,\bar\eta\right)$ depends
strongly on the treatment of the errors.  We use the following
procedure: first we apply \eq{epsKcond} to find the CKM phase $\delta$
of the standard parametrization from the input parameters, which are
scanned in an 1$\sigma$ ellipsoid of their errors.  Second, we check
the consistency of the obtained phases $\delta$ with \bbmd
{\eq{condBBmd}}.  Here we treat the errors in are fully conservative
way.  Last we apply the constraint from lower limit on $\DmBs$
{\eq{condBBms}}.  This constraint is very sensitive to the value of
the flavour-SU(3) breaking term $\fbs\sqrt{\bbs}/\fbd\sqrt{\bbd}$.
Using the quenched lattice QCD value \eq{xisd} one finds the allowed
values of $\left(\bar\rho,\bar\eta\right)$ as displayed in
{\fig{fig:dmbs-ut}}.
\begin{nfigure}
\includegraphics[clip,width=\miniwidth]{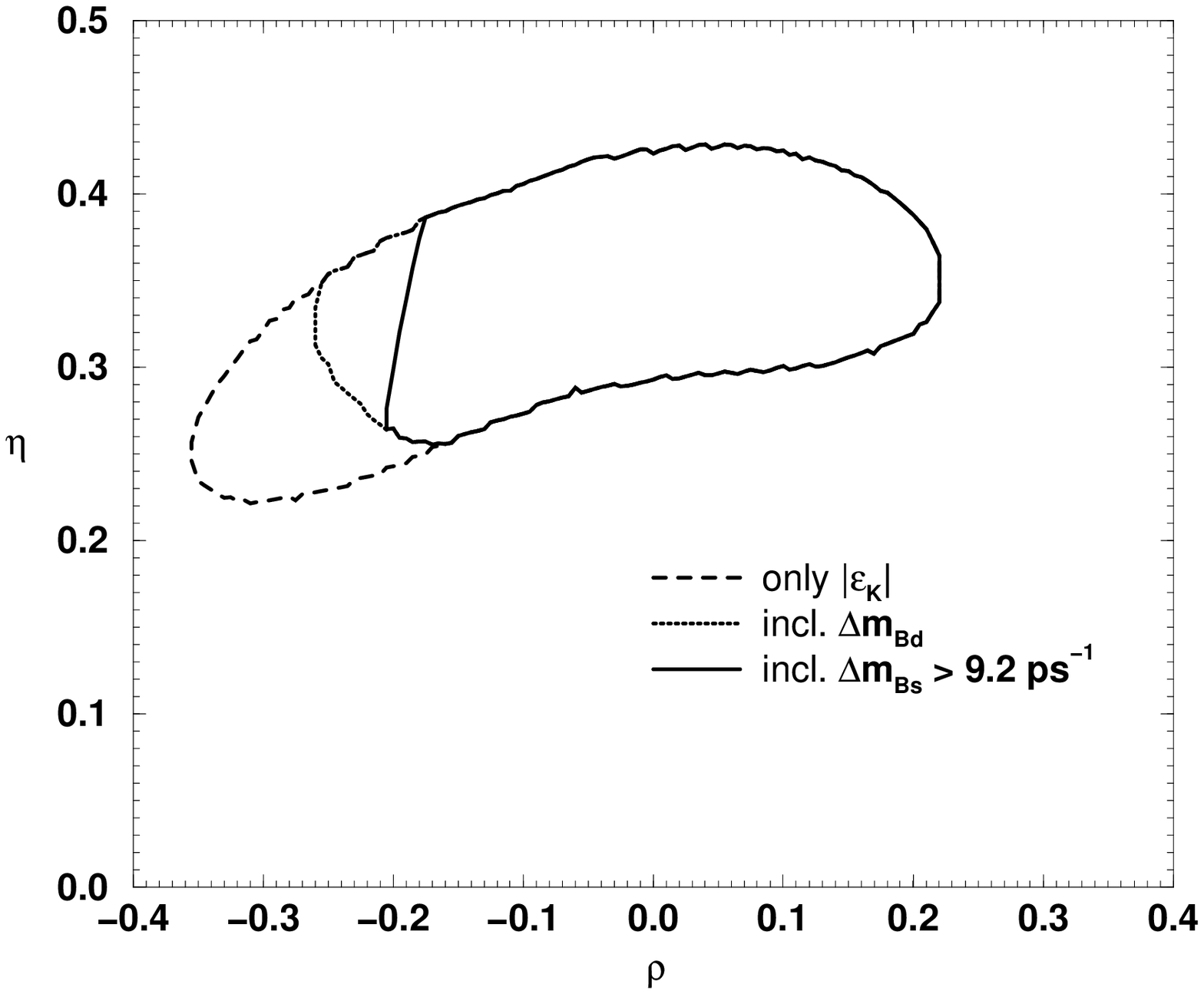}
\ncaption{The allowed values of $\left(\bar\rho,\bar\eta\right)$.  The
outer contour is obtained solely from $\left|\epsK\right|$ and
unitarity, the medium one takes into account \bbmd, the inner curve
\bbms\ \eq{LimitDmBs} using the quenched lattice value \eq{xisd} for
illustrative reasons.  If one would use a 10\% higher value for the
flavour SU(3) breaking as expected for an unquenched calculation no
effect is visible for the current limit \eq{LimitDmBs}.}
\label{fig:dmbs-ut}
\end{nfigure}
If one would increase $\fbs\sqrt{\bbs}/\fbd\sqrt{\bbd}$ by 10\% as
expected for an unquenched calculation, no effect is visible for the
current limit \eq{LimitDmBs}.  This can be read off from
{\fig{fig:dmbs-ratio}}, where we plot the fraction of area cut out
from the allowed region of $\left(\bar\rho,\bar\eta\right)$ by the
$\DmBs$ constraint as a function of $\DmBs$.
\begin{nfigure}
\includegraphics[clip,width=\miniwidth]{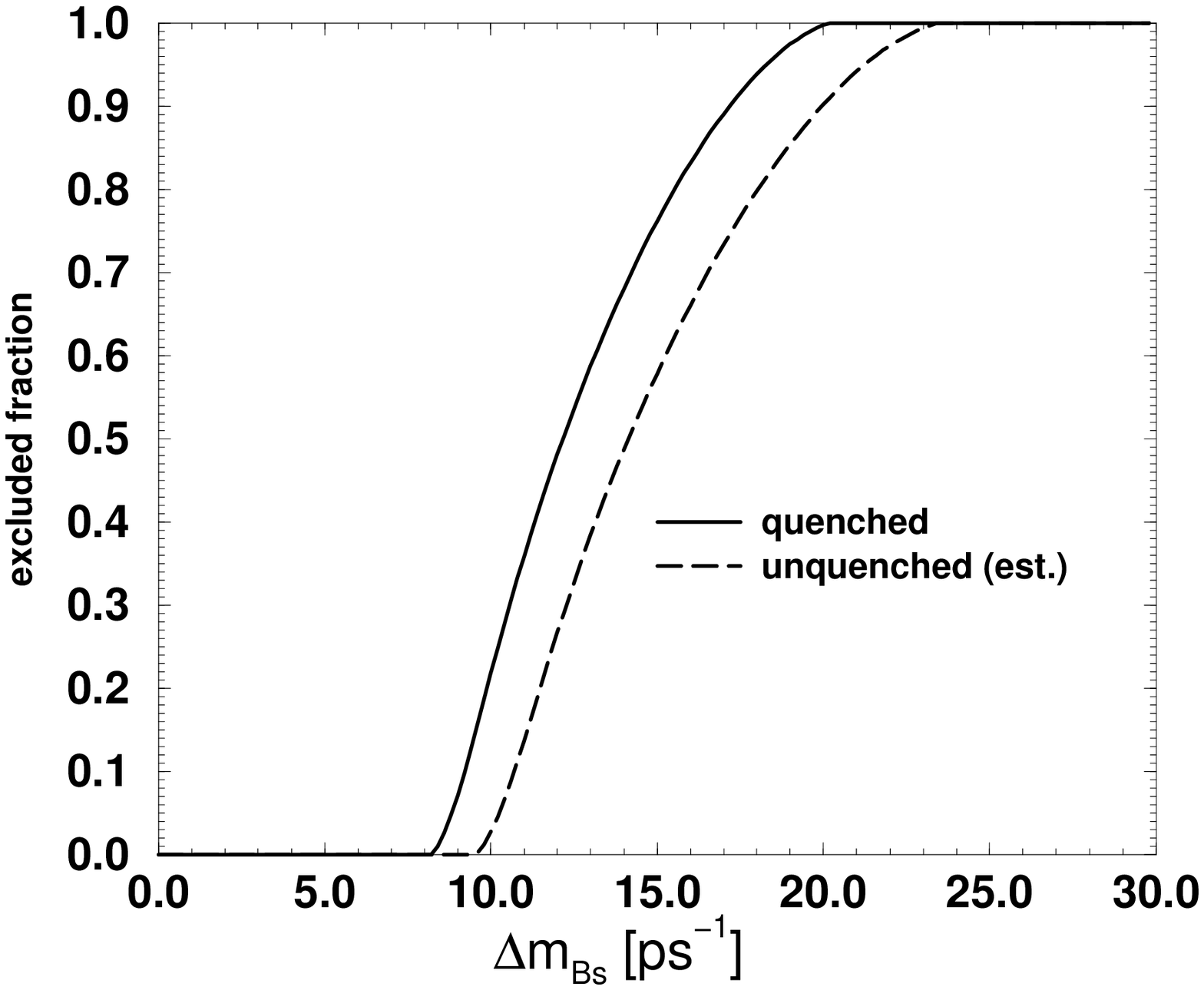}
\ncaption{ The fraction of the allowed area for
$\left(\bar\rho,\bar\eta\right)$ which is excluded by the constraint
from \bbms\ as a function of $\DmBs$.  The curve labelled ``quenched''
is obtained using \eq{xisd}, the line labelled ``unquenched (est.)''
uses a 10\% larger value.}
\label{fig:dmbs-ratio}
\end{nfigure}

From \fig{fig:dmbs-ut} we read off the allowed ranges of the
parameters describing the unitarity triangle:
\begin{equation}
\begin{array}{c}
40^\circ \leq \alpha \leq 101^\circ, \quad
57^\circ \leq \gamma \leq 127^\circ, \\
0.42 \leq \sin\!\left(2\beta\right) \leq 0.79 \\
-0.20 \leq \bar\rho \leq 0.22, \quad
0.25 \leq \bar\eta \leq 0.43.
\end{array}
\end{equation}

\section*{Acknowledgements}
I would like to thank Guido Martinelli for a clarifying discussion on
$\bk$ at this conference.

% make this last stuff fit on this page
\enlargethispage*{2ex}

\end{document}